\documentclass[aps, reprint, twocolumn,superscriptaddress,longbibliography]{revtex4-1}

\usepackage[latin1]{inputenc}
\usepackage{graphicx}
\usepackage{amsmath,amssymb,amstext,graphicx,bm,booktabs,hyperref,siunitx,upgreek}
\usepackage{physics}

\newcommand{\natcs}{\ensuremath{\mathrm{N@C}_{60}}}
\newcommand{\fnatcs}{\ensuremath{^{15}\mathrm{N@C}_{60}}}
\newcommand{\cs}{$\mathrm{C}_{60}$}
\newcommand{\cstwo}{$\mathrm{CS}_{2}$}

\begin{document}

\title{The spin resonance clock transition of the endohedral fullerene $^{\mathbf 15}$N@C$_\mathbf{60}$}

\author{R.T.~Harding}
\affiliation{Department of Materials, University of Oxford, Parks Road, Oxford OX1 3PH, United Kingdom}

\author{S.~Zhou}
\affiliation{Department of Materials, University of Oxford, Parks Road, Oxford OX1 3PH, United Kingdom}

\author{J.~Zhou}
\affiliation{Department of Materials, University of Oxford, Parks Road, Oxford OX1 3PH, United Kingdom}

\author{T.~Lindvall}
\affiliation{VTT Technical Research Centre of Finland Ltd, Centre for Metrology MIKES, P.O. Box 1000, FI-02044 VTT, Finland}

\author{W.K.~Myers}
\affiliation{Centre for Advanced Electron Spin Resonance, Inorganic Chemistry Laboratory, University of Oxford, South Parks Road, Oxford OX1 3QR, United Kingdom}

\author{A.~Ardavan}
\affiliation{Clarendon Laboratory, Department of Physics, University of Oxford, Parks Road, Oxford OX1 3PU, United Kingdom}

\author{G.A.D.~Briggs}
\affiliation{Department of Materials, University of Oxford, Parks Road, Oxford OX1 3PH, United Kingdom}

\author{K.~Porfyrakis}
\affiliation{Department of Materials, University of Oxford, Parks Road, Oxford OX1 3PH, United Kingdom}

\author{E.A.~Laird}
\affiliation{Department of Materials, University of Oxford, Parks Road, Oxford OX1 3PH, United Kingdom}

\begin{abstract}

The endohedral fullerene $^{15}\mathrm{N@C}_{60}$ has narrow electron paramagnetic resonance lines which have been proposed as the basis for a condensed-matter portable atomic clock. We measure the low-frequency spectrum of this molecule, identifying and characterizing a clock transition at which the frequency becomes insensitive to magnetic field. We infer a linewidth at the clock field of 100 kHz. Using experimental data, we are able to place a bound on the clock's projected frequency stability. We discuss ways to improve the frequency stability to be competitive with existing miniature clocks.
\end{abstract}

\date{\today{}}
\maketitle

Precise and portable frequency references underpin a range of infrastructure for navigation, communication, and sensing~\cite{Vig1993,Shkel2010}. The state-of-the-art miniaturised frequency standard is the chip-scale atomic clock~\cite{Knappe2004}, which uses optical interrogation of alkali metal vapor \cite{Cyr1993}. However, a condensed-matter clock could further improve size, weight, power, and cost by obviating the need for optical and vacuum elements.
One approach is to exploit a radio-frequency spin resonance transition as a reference, as proposed using V$^{++}$ impurities in MgO~\cite{White2005} or nitrogen vacancy centres in diamond~\cite{Hodges2013}. In these approaches, it is essential to operate at a transition where the resonance is not detuned by magnetic field fluctuations.

Here, we detect and characterize a field-independent `clock transition' in the endohedral fullerene molecule \fnatcs, demonstrating its suitability as a frequency reference. This molecule is a promising candidate for a condensed-matter atomic clock~\cite{Hannah2007,briggs2012atomic} because of its sharp resonances~\cite{Morton2006} and the potential for circuit integration using the approach of ``1-chip'' nuclear magnetic resonance \cite{Sun2011,Anders201641,Grisi2015}. Such an atomic clock would lock the frequency of a local oscillator to the reference provided by the spin resonance. We measure the clock transition frequency and field, place an upper limit on the linewidth, measure the signal strength, and set out the requirements to use this molecule as a portable frequency standard. 

The endohedral fullerene \fnatcs~comprises a nitrogen atom encapsulated within a carbon cage \cite{PhysRevLett.77.1075}. The nitrogen, located at the cage centre, retains its ground state electronic configuration $^4S_{3/2}$ \cite{PSSA:PSSA81}, with spin quantum numbers $S=3/2$ for the electron and $I=1/2$ for the nucleus. The energy levels of this molecule in a magnetic field $\boldsymbol{B_0}=B_0\,\hat{\boldsymbol{z}}$ are described by the Hamiltonian
\begin{equation}\label{Equation1}
\mathcal{H}=g_\mathrm{e} \mu_\mathrm{B} S_z B_0 - g_\mathrm{N} \mu_\mathrm{N} I_z B_0 + A\hat{\boldsymbol{S}}\cdot\hat{\boldsymbol{I}},
\end{equation}
where $g_\mathrm{e}$ $(g_\mathrm{N})$  is the $g$-factor of the electron (nucleus), $\mu_\mathrm{B}$ $(\mu_\mathrm{N})$ is the Bohr (nuclear) magneton and $A$ is the isotropic hyperfine constant \cite{Morton2006}. The electron and nuclear spin operators are denoted by $\hat{\boldsymbol{S}}$ and $\hat{\boldsymbol{I}}$ respectively. Zero-field splitting is neglected due to the near-spherical symmetry of the system \cite{thesis:Morley}.

\begin{figure}
\includegraphics[width=\columnwidth]{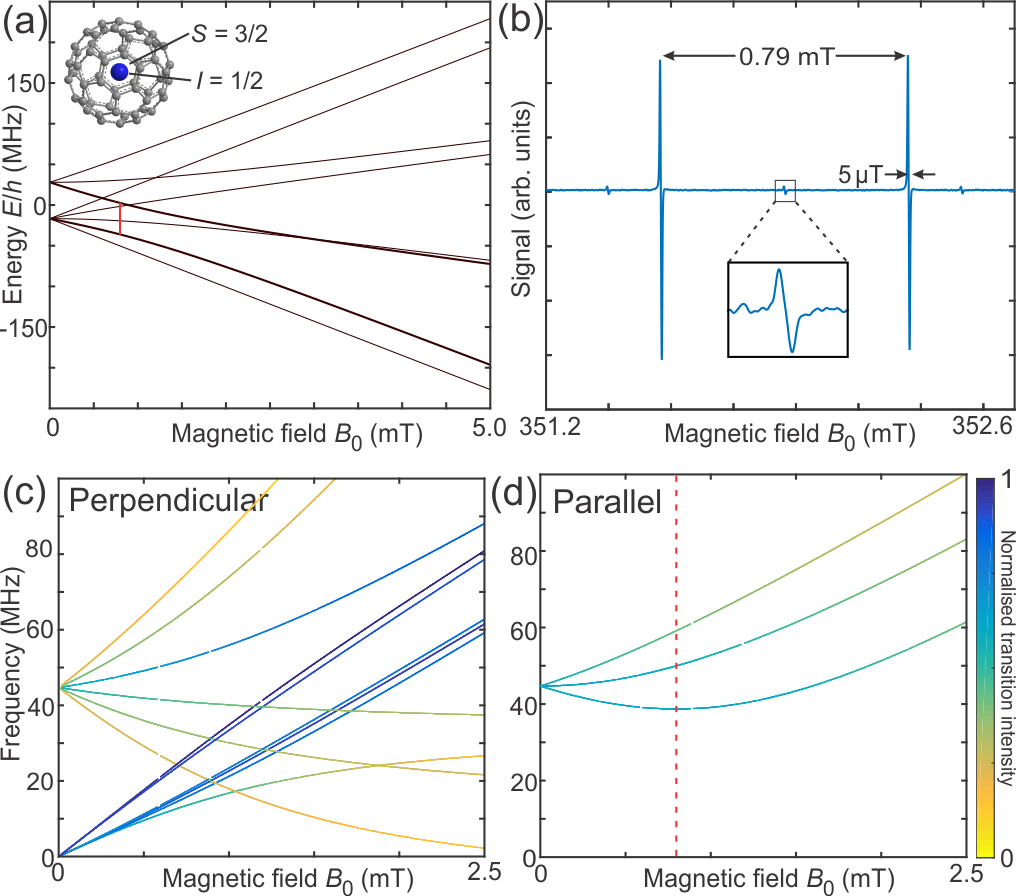}
\caption{(a) Simulated energy levels of \fnatcs~as a function of magnetic field. The clock transition is denoted by a vertical red line. Inset: molecular structure, labelled with electron ($S$) and nuclear ($I$) spins. (b) EPR spectrum of \fnatcs~measured at X-band ($f=9.86\,\si{\giga\hertz}$). The zoom-in highlights the contribution of residual $^{14}$\natcs. (c) and (d): Simulated frequencies of allowed transitions for driving magnetic field applied perpendicular (c) and parallel (d) to the static magnetic field $B_0$. Traces are colored according to transition intensity. The clock field is marked with a dashed line.
\label{Figure1}
}
\end{figure}

Figure~\ref{Figure1}(a) shows simulated energy eigenstates of  Hamiltonian~(\ref{Equation1}) as a function of magnetic field. The simulation parameters are $g_\mathrm{e}=2.00204$ and $|A|/h=22.35\,\si{\mega\hertz}$ \cite{PIETZAK1998613}, with $g_\mathrm{N}=-0.566$~\cite{Baldeschwieler1962,STONE200575}. The negative nuclear $g$-factor leads to a negative value of $A$, leading to an inverted energy level manifold \cite{vanier1989quantum}. Near zero field, the eigenstates form a pair of hyperfine-split multiplets labelled by quantum numbers $\ket{F,m_F}$, where $\hat{\boldsymbol{F}}=\hat{\boldsymbol{I}}+\hat{\boldsymbol{S}}$.
At high fields, where the Zeeman interaction dominates, the level diagram simplifies to four doublets labelled by $\ket{m_S,m_I}$.
Between these limits lies a crossover region in which the eigenstates are mixed in either basis.

In a magnetic resonance experiment, the allowed transitions between energy levels depend on the orientation of the oscillating drive field $\boldsymbol{B_1}$ relative to $\boldsymbol{B_0}$. 
The frequencies of allowed transitions are plotted in Fig.~\ref{Figure1}(c) and (d) for perpendicular mode ($\boldsymbol{B_1} \perp \boldsymbol{B_0}$) and parallel mode ($\boldsymbol{B_1} || \boldsymbol{B_0}$).
To indicate the strength of each transition, traces are plotted with a color scale corresponding to the magnitude of the transition matrix element $\bra{f}g_\mathrm{e}S_{x(z)}+g_\mathrm{N}I_{x(z)}\ket{i}$ for perpendicular- \mbox{(parallel-)} mode transitions between initial state $\ket{i}$ and final state~$\ket{f}$.

The clock transition occurs at an anticrossing between two energy levels that are mixed by the hyperfine interaction (Fig.~\ref{Figure1}(a)). At the clock field, the transition frequency is insensitive to magnetic field ($\mathrm{d}f/\mathrm{d}B=0$). This transition can be driven in parallel mode (Fig.~\ref{Figure1}(d)) and occurs at a field $B_{\mathrm{clock}}=|A|/(g_\mathrm{e} \mu_\mathrm{B} + g_\mathrm{N} \mu_\mathrm{N}) \approx 0.8\,\si{\milli\tesla}$ with a frequency $f_{\mathrm{clock}}=\sqrt{3}|A|/h\approx 38$~MHz.

\begin{figure}
	\includegraphics[width=\columnwidth]{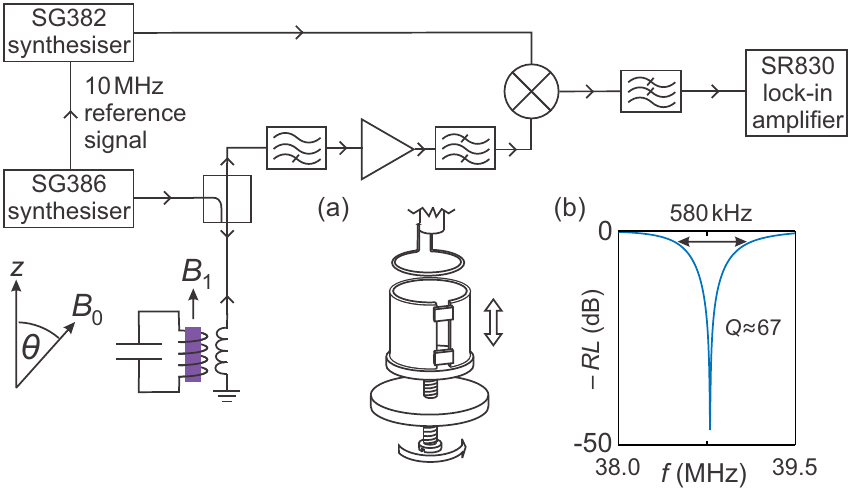}
\caption{Measurement setup. An excitation source and local oscillator are implemented using two phase-locked signal generators. The signal reflected from the resonator is high-pass filtered before amplification, in order to suppress inductive pickup of the modulation field. Inset (a): loop-gap resonator arrangement with adjustable coupling between antenna and resonator. Inset (b): Reflection from the resonator, demonstrating near-critical coupling and good return loss $RL$.
		\label{Figure2}
	}
\end{figure}

An \fnatcs~sample was prepared by ion implantation of $\mathrm{C}_{60}$ using $^{15}\mathrm{N}_2$ (isotopic purity $98\%$)~\cite{PhysRevLett.77.1075}. The crude sample was dissolved in toluene and purified by high performance liquid chromatography to a concentration (\fnatcs:\cs) of $\sim$6000\,ppm~\cite{Kanai2004}. The sample was characterized at X-band (Fig.~\ref{Figure1}(b)), showing a pair of narrow spin resonances  as expected from hyperfine interaction in \fnatcs~\cite{PhysRevLett.77.1075}. For low-field experiments, the sample was redissolved in deoxygenated \cstwo~under $\mathrm{N}_2$ atmosphere ($<0.01\,\mathrm{ppm}$ $\mathrm{O}_2$)~\cite{Morton2007} and flame sealed in a Suprasil tube. The estimated spin number is $N_{\mathrm{spin}}\approx1.4\times10^{16}$ in a volume $0.55\,\si{\centi\metre}^3$.

To characterize the clock transition, we performed fixed-frequency, swept-field continuous-wave spin resonance measurements using the circuit of Fig.~\ref{Figure2}. 
The swept magnetic field $\boldsymbol{B}_\mathrm{coil}$ was generated using an air-cored Helmholtz electromagnet oriented to select either perpendicular or parallel mode operation. Small stray magnetic fields shift $\boldsymbol{B}_\mathrm{0}$ away from $\boldsymbol{B}_\mathrm{coil}$.
To obtain each spectrum, the sample was loaded into one of a series of interchangeable loop-gap resonators (LGRs)~\cite{FRONCISZ1982515} spanning the range 38--70$\,\si{\mega\hertz}$. The resonant frequency of each LGR was adjusted using high-$Q$ non-magnetic chip capacitors soldered across the gap~\footnote{0505C series non-magnetic  high-Q, low ESR (NP0 TC) capacitors, Passive Plus, Inc., New York, USA}. 

The resonance signal was detected using radio-frequency (RF) reflectometry. The LGR was probed via a coupling antenna, with the separation adjusted mechanically to achieve near-critical coupling (return loss $RL\!>\!50$\,dB and loaded quality factor $Q\sim 70$). An RF excitation, with power $P=-21\,\mathrm{dBm}$ chosen for good signal-to-noise ratio, was applied to the LGR via a directional coupler. The reflected signal was filtered and amplified before being mixed with the local oscillator signal to generate a homodyne voltage.
The phase between the local oscillator and the excitation source was chosen to maximise the dc component of this voltage, which maximises sensitivity to the amplitude of the reflected signal. The excitation frequency was set 3~kHz above the LGR frequency to suppress the effects of drift in the coupling \cite{book:Poole1983}. 
For improved sensitivity, a small modulation field $\boldsymbol{B}_{\mathrm{mod}}$ was applied parallel to $\boldsymbol{B}_0$ at a frequency of 5319\,Hz, allowing the homodyne voltage to be be detected using a lock-in amplifier synchronized with this modulation.
The modulation amplitude was measured using a search coil as $B_{\mathrm{mod}}\approx 1.1\times10^{-5}\,\si{\tesla}$.

The oscillating magnetic field $\boldsymbol{B_1}$ generated by the RF excitation is linearly polarised along the axis of the LGR. 
Its amplitude is $B_1\approx\sqrt{\frac{\mu_0 P Q}{2\pi Vf}} \approx 1.8\,\si{\micro\tesla}$, where $\mu_0$ is the permeability of free space, $V\approx 0.8$~cm$^3$ is the resonator volume, and $f$ is the frequency. This equation is derived assuming perfect impedance matching, a spatially homogeneous RF field within the resonator, and the equal division of magnetic field energy between the volume enclosed by the resonator and the return flux path.

\begin{figure}
\includegraphics[width=\columnwidth]{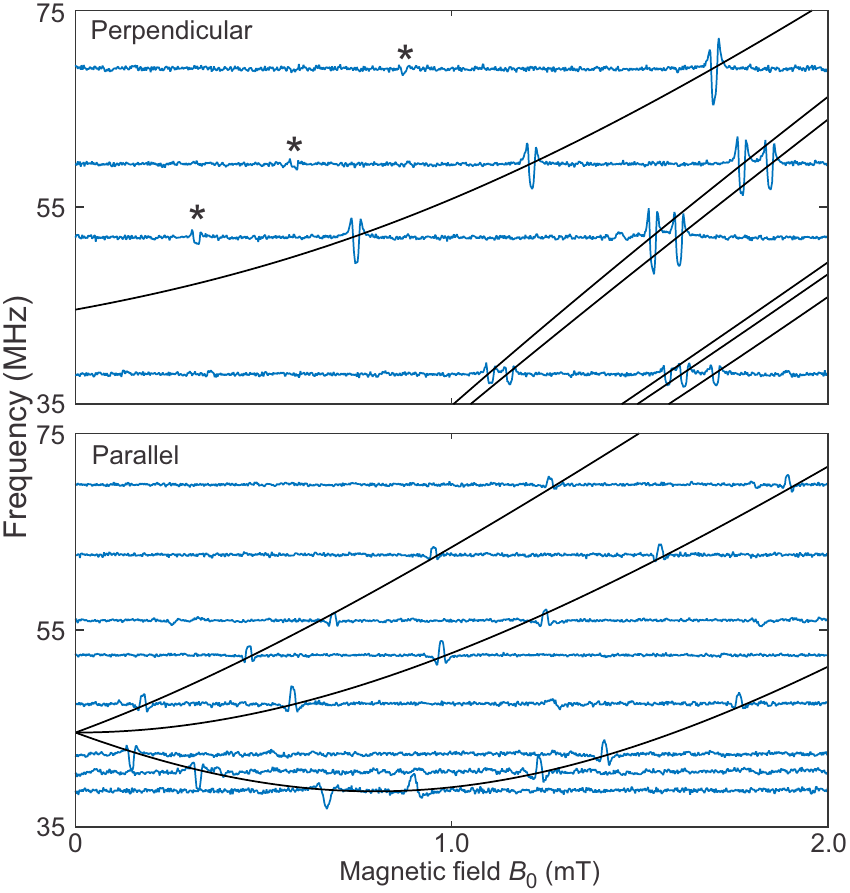}
\caption{Perpendicular- and parallel-mode EPR spectrum of \fnatcs~at selected excitation frequencies. Blue: Lock-in signal (arbitrary units) as a function of rescaled magnetic field $B_0$, offset such that the baseline of each trace aligns with the corresponding excitation frequency. Each trace is an average of up to 100 sweeps. Black: fit to resonant fields using exact solutions to Hamiltonian~(\ref{Equation1}). In perpendicular mode, it is possible to observe resonances (marked by $\star$) that were not fitted due to low signal-to-noise ratio.
\label{Figure3}
}
\end{figure}

The low-field EPR spectra are shown in Fig.~\ref{Figure3}, and match well the predictions of Fig.~\ref{Figure1}(c-d). The resonant fields are extracted by fitting the peaks with the second derivative of a Lorentzian.
For quantitative analysis, we fit these resonant fields using the theoretical values calculated from Eq.~(\ref{Equation1})~\cite{supp}. We account for small differences between the nominal field $B_\mathrm{coil}$ and the true field $B_0$ at the sample by fitting the data using $B_0=\alpha B_\mathrm{coil} + B_\mathrm{offset}$, where $\alpha$ is a correction parameter, reflecting sample misalignment and uncertainty in the calculated coil constant, and $B_\mathrm{offset}$ is an environmental offset field. Fitting parallel and perpendicular datasets simultaneously using $A$, $\alpha$, and $B_\mathrm{offset}$ as fit parameters, while fixing $g_\mathrm{e}=2.00204$ and $g_\mathrm{N}=-0.566$, we extract $|A|/h= 22.30\pm0.02\,\si{\mega\hertz}$, $\alpha=1.024\pm0.003$, and $B_\mathrm{offset}=-28\pm4\,\si{\micro\tesla}$, where error bars are one standard deviation intervals derived from the fits. The extracted $\alpha$ and $B_\mathrm{offset}$ are used to plot the data in terms of $B_0$ rather than $B_\mathrm{coil}$.

\begin{figure}
	\includegraphics[width=\columnwidth]{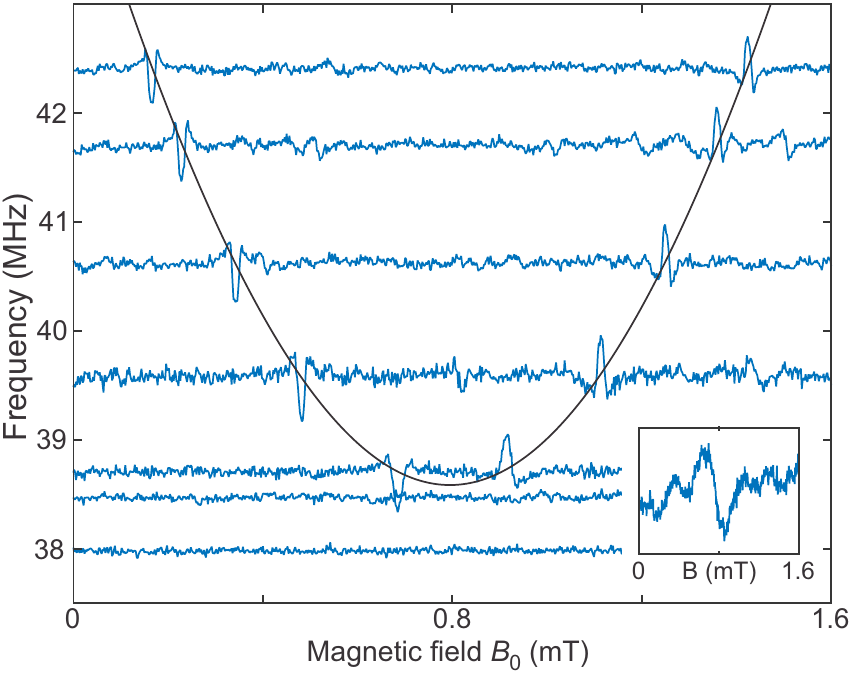}
	\caption{Parallel-mode EPR spectrum of the clock transition near the clock field at selected excitation frequencies. Blue: Lock-in signal (arbitrary units) as a function of rescaled magnetic field at each frequency. Black line: fit to resonant fields. Inset: spectrum measured at $38.474\,\si{\mega\hertz}$, cutting near the tip of the parabola. A larger modulation amplitude was necessary to detect the nearly field-independent transition.
		\label{Figure4}
	}
\end{figure}

To characterize the clock transition in detail, Fig.~\ref{Figure4} shows the parallel-mode spectra near the clock field. Fitting these data as in Fig.~\ref{Figure3}, we extract $|A|/h= 22.277\pm 0.001\,\si{\mega\hertz}$,  $\alpha=1.0257\pm0.0005$, and $B_\mathrm{offset}=-12.1\pm0.5\,\si{\micro\tesla}$. The value of $A$ is consistent between Fig.~\ref{Figure3} and Fig.~\ref{Figure4}, and lies within the range of previous X-band measurements \cite{PSSB:PSSB453,PIETZAK1998613}. The values obtained for $B_\mathrm{offset}$ are much smaller than $B_\mathrm{coil}$, and are consistent with the geomagnetic field.

\begin{figure} 
\includegraphics[width=\columnwidth]{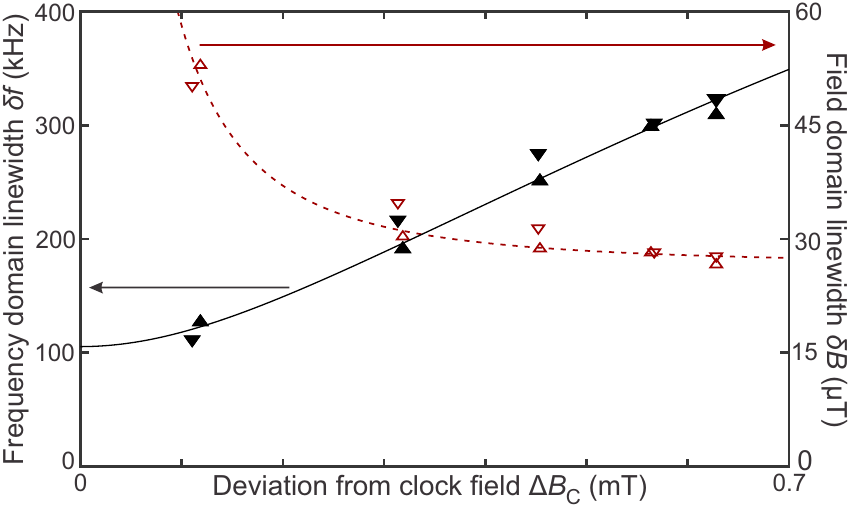}
\caption{Linewidth as a function of deviation from the clock field. Left axis: frequency-domain linewidth $\delta f$ above ($\blacktriangle$) and below ($\blacktriangledown$) the clock field. Right axis: field-domain linewidth $\delta B$ above ($\pmb{\bigtriangleup}$) and below ($\pmb{\bigtriangledown}$) the clock field. Solid black line: fit to the frequency-domain linewidth generated using Eq.~(\ref{Equation_frequency_fit}). Dashed red line: the same fit mapped to the field domain.
\label{Figure5}
}
\end{figure}

Measuring near the clock field reduces the sensitivity of the transition frequency to magnetic field fluctuations. To quantify this, we extract field-domain linewidths $\delta B$ from the spectra of Fig.~\ref{Figure4}~\cite{supp}. As a function of the deviation from the clock field~$\Delta B_{\mathrm{C}}=|B_0 - B_\mathrm{clock}|$, the field-domain linewidth $\delta B$ increases towards the clock field~(Fig.~\ref{Figure5}). However, when this field linewidth is converted to a frequency-domain linewidth $\delta f = \left| \frac{\mathrm{d}f}{\mathrm{d}B_0}\right| \delta B$, the resonance sharpens toward the clock transition as expected.
Although our field-swept configuration precludes directly measuring the linewidth at the clock field, we are able to estimate the corresponding dephasing time $T_2^*$ using a simple model in which an intrinsic linewidth set by $T_2^*$ adds in quadrature with broadening due to magnetic field fluctuations of strength $B_\mathrm{var}$:
\begin{equation}
\left(\delta f (\Delta B_\mathrm{C})\right)^2=\left(\frac{1}{\pi T_2^*}\right)^2+\left(\frac{\mathrm{d}f (\Delta B_\mathrm{C})}{\mathrm{d}B}\cdot B_\mathrm{var}\right)^2.
\label{Equation_frequency_fit}
\end{equation}
Here $B_\mathrm{var}$ accounts for line-broadening due to both the modulation field and environmental field noise. Fitting with this model (Fig.~\ref{Figure5}) gives $T_2^*=3.0\pm0.4\,\si{\micro\second}$ and $B_\mathrm{var}=26.2\pm0.9\,\si{\micro\tesla}$, including the contribution of the modulation field.

These measurements enable an assessment of the feasibility of an endohedral fullerene frequency standard. The Allan deviation $\sigma_y~(\uptau)$, which parameterises the fractional frequency stability over time $\uptau$~\cite{1539968}, can be estimated as~\cite{riehle_frequency_standards}:
\begin{equation}
\sigma_y(\uptau)\approx \frac{1}{\mathrm{SNR}\cdot Q_A}\sqrt{\frac{1}{\uptau}},
\label{eq:sigmay}
\end{equation}
where $\mathrm{SNR}$ and $Q_A$ are the signal-to-noise ratio and quality factor of the clock transition resonance signal respectively. The signal-to-noise ratio is $\mathrm{SNR}=S_0/\delta S$, where $S_0$ is the signal amplitude and $\delta S$ is the root mean square noise per $\sqrt{\si{\hertz}}$ bandwidth; the transition quality factor is defined by $Q_A\equiv f_\mathrm{clock}/\delta f$. Equation (\ref{eq:sigmay}) applies for short measurement times $\uptau$, such that the noise spectrum is approximately white. Considering our experimentally measured parameters, with signal amplitude $S_0 \approx 5\,\si{\nano\volt}$, input-referred noise voltage $\delta S \approx 2.5\,\si{\nano\volt\,\hertz^{-1/2}}$, clock frequency $f_\mathrm{clock}=\sqrt{3}|A|\approx38.6\,\si{\mega\hertz}$, and resonance linewidth $\delta f \approx 100\,\si{\kilo\hertz}$, Eq.~(\ref{eq:sigmay}) predicts a short-term Allan deviation $\sigma_y \approx 1.3\times10^{-3}\,\uptau^{-1/2}$ for a \fnatcs-based atomic frequency standard.

To improve the performance, it will be necessary to optimise both $\mathrm{SNR}$ and $Q_A$. Improvements to the signal strength would come from increased spin density, achieved by using higher-purity material and by choosing a solvent that allows for a higher molecular concentration, or even a powder sample. Further improvements to $\mathrm{SNR}$ could be achieved with a larger sample volume to increase signal, or by reducing electronic noise. To increase $Q_A$, the linewidth should be reduced below our measured value, the origin of which is currently unknown. For comparison, coherence time $T_2 = 80\,\si{\micro\second}$ was attained in a carefully prepared $^{14}$\natcs~sample measured at X-band~\cite{Morton2006}, a significant improvement over the $T_2^*$ value measured here. Finally, using a molecular species engineered for a higher clock frequency will improve both $Q_A$ and SNR. As an example, solid-state molecular systems with a clock frequency $f_\mathrm{clock}\sim 10\,\si{\giga\hertz}$ have been demonstrated \cite{Shiddiq2016}.

\label{Section4}

We now consider what would be necessary to make an endohedral fullerene clock competitive with existing vapor-based miniature atomic clocks, which achieve short-term Allan deviations $\sigma_y \leq 3 \times 10^{-10}$ for $\uptau=1\,\si{\second}$~\cite{csac_datasheet}.
The spin density in our experiment ($n \approx 2.5 \times 10^{22}~\mathrm{m}^{-3}$) approaches the limit set by the saturation concentration of $\mathrm{C}_{60}$ in CS$_2$ ($7.9\,\si{\milli\gram\per\milli\litre}$) \cite{Ruoff1993} given the available sample purity.
Using pure endohedral fullerene material dissolved in \mbox{1-chloronaphthalene} (saturation concentration $51\,\si{\milli\gram\per\milli\litre}$) would give $n \approx 4.3 \times 10^{25}~\mathrm{m}^{-3}$. To further improve the signal, the LGR could be replaced by a solenoid with a similar filling factor and improved RF quality factor $Q\approx200$~\cite{Hoult1976}, leading to a roughly three-fold increase in signal. Additionally, there is scope to reduce the voltage noise in our setup from $\delta S\approx 2.5\,\si{\nano\volt\,\hertz^{-1/2}}$ to the room-temperature Johnson-Nyquist limit $\delta S\approx 0.9\,\si{\nano\volt\,\hertz^{-1/2}}$. Combining all these improvements while maintaining the  sample volume would lead to an increase in SNR by approximately a factor of $1.4\times 10^4$, leading to a short-term Allan deviation $\sigma_y \sim 1 \times 10^{-7} \uptau^{-1/2}$ given our measured linewidth. If the clock transition linewidth could be reduced to $\delta f \sim 100\,\si{\hertz}$, a demanding requirement, the Allan deviation would then be $\sigma_y \sim 1\times 10^{-10}\,\uptau^{-1/2}$ for an atomic frequency standard based on \fnatcs. Alternatively, using a powder sample ($n \approx 1.4 \times 10^{27}~\mathrm{m}^{-3}$) instead of a solution would allow the same performance to be achieved with the larger linewidth $\delta f \sim4\,\si{\kilo\hertz}$. However, in both powder and high-density solution, it is critical that dipolar coupling does not broaden these linewidths. Such broadening is typically suppressed at the clock field~\cite{Wolfowicz2013, Shiddiq2016}.

For large $\uptau$, $\sigma_y (\uptau)$ is greater than the prediction of Eq.~\eqref{eq:sigmay} because of drifts in the clock parameters. For example, thermal drift enters via the temperature dependence of the hyperfine constant~\cite{book:endofullerenes}, known from $^{14}$\natcs~measurements to be $\frac{1}{A}\frac{dA}{dT} = 89$~ppm/K. To mitigate this effect, it will be necessary to precisely monitor the temperature, for example by measuring simultaneously three transitions, which together allow errors in frequency, magnetic field, and temperature to be extracted.

As a route to improving both short- and long-term stability, the endohedral fullerene $^{31}\mathrm{P@C}_{60}$, with the same spin quantum numbers as \fnatcs but larger hyperfine coupling, offers a clock frequency $f_\mathrm{clock} \approx 240\,\si{\mega\hertz}$ \cite{Knapp1998}. Given a linewidth $\delta f \approx 350\,\si{\kilo\hertz}$ \cite{Knapp1998}, the predicted transition quality factor $Q_A\approx700$ for $^{31}\mathrm{P@C}_{60}$ compares favourably to the measured value $Q_A\approx400$ for \fnatcs.  The $^{31}\mathrm{P@C}_{60}$-based frequency standard would also benefit from an approximately twenty-fold improvement due to the frequency-dependent $\mathrm{SNR}$ scaling, where $\mathrm{SNR}\propto f_\mathrm{clock}^{7/4}$ \cite{Hoult1976}. Assuming the same sample volume, concentration and purity as considered above, the improved SNR and $Q_A$ would lead to a forty-fold improvement of the predicted short-term Allan deviation relative to that of the \fnatcs-based frequency standard, such that $\sigma_y \sim 3\times10^{-9}\,\uptau^{-1/2}$ might be achieved in solution without further improvements in linewidth. To achieve $\sigma_y \sim 1\times 10^{-10}$ would require a linewidth $\delta f \sim 10\,\si{\kilo\hertz}$. Furthermore, $^{31}\mathrm{P@C}_{60}$ exhibits a weaker temperature dependence $\frac{1}{A}\frac{dA}{dT} = 74$~ppm/K~\cite{book:endofullerenes}, reducing the sensitivity to temperature fluctuations.

We have demonstrated that \fnatcs~possesses a spin resonance clock transition and therefore fulfils one of the criteria for a condensed matter atomic clock. This is the first observation of a clock transition in a molecular system at room temperature. The data presented here do not yet demonstrate clock operation, which would require a swept-frequency measurement at the clock field. Furthermore, useful operation will require purer material and a much narrower linewidth. For a chip-scale clock, the spectrometer  also needs to be miniaturised.
Our approach would eliminate the complexity of vapor-based clocks in favor of a single radio-frequency circuit. This would enable further miniaturization, robustness, and ease of manufacture, leading to a wider range of application of atomic clocks.

We thank J.\ Liu, J.\ F.\ Gregg, P.~Dallas, and C.\ J.\  Wedge for useful discussions, and J.\ J.\ Le~Roy for assistance with sample preparation. We acknowledge DSTL, EPSRC (EP/J015067/1, EP/K030108/1, EP/N014995/1, EP/P511377/1), the Royal Academy of Engineering, a Marie Curie CIG award, and LocatorX Inc. of Jackson Beach, Florida.

%


\clearpage
	\setcounter{equation}{0}
	\setcounter{figure}{0}
	\setcounter{table}{0}
	\setcounter{page}{1}
	\makeatletter
	\renewcommand{\theequation}{S\arabic{equation}}
	\renewcommand{\thefigure}{S\arabic{figure}}
	\renewcommand{\thetable}{S\Roman{table}}
	\renewcommand{\bibnumfmt}[1]{[S#1]}
	\renewcommand{\citenumfont}[1]{S#1}
\parbox[c]{\textwidth}{\protect \centering \Large \MakeUppercase{Supplementary Material}}
\rule{\textwidth}{1pt}

\section{Lineshape}
Here, we derive the lineshape used to fit the resonances displayed in Figure 3 and Figure 4 of the main text, which differs from the lineshape typically observed in continuous-wave electron paramagnetic resonance (EPR) experiments. In most EPR experiments, the spectrometer is adjusted to maximise sensitivity to the absorption of radiation in the spin ensemble rather than its effect on the phase of radiation reflected from the resonator. This is achieved by using automatic frequency control (AFC) to lock the frequency of the excitation to that of the resonator, and selecting a suitable local oscillator phase, in order to suppress the response due to the component of magnetic susceptibility that corresponds to phase shifts. When used in conjunction with field- or frequency-modulation and lock-in detection, the observed signals are therefore described by the first derivative of an absorption lineshape, as shown in Figure 1(b) of the main text.

In contrast, the signals displayed in Figure 3 and Figure 4 are well fitted by the second derivative of a Lorentzian. This second-derivative lineshape is typical for signals due to the phase component observed using lock-in detection. In this experiment, the spectrometer is sensitive to the phase response of the spin ensemble despite the phase of the local oscillator being chosen to maximise sensitivity to changes in amplitude of radiation reflected from the LGR. This is because the excitation frequency was slightly detuned from the cavity resonant frequency and, in the absence of AFC, the shift in the resonator frequency caused by the spin ensemble is converted to an amplitude signal by the slope of the resonator dip.

\begin{figure}[h!]
\centering
\includegraphics{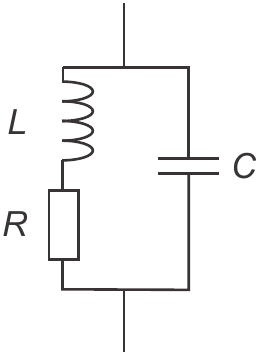}
	\caption{Lumped element model of loop gap resonator, comprising inductance $L$ (loop), capacitance $C$, and resistance $R$.}\label{sup_fig:resonator}
\end{figure}

To understand this phenomenon quantitatively, we consider the effect of the spin ensemble on the LGR resonant frequency  $\omega_\mathrm{res}$. The LGR is modelled using the circuit displayed in Figure \ref{sup_fig:resonator}, which comprises an inductance $L$, capacitance $C$, and a resistance $R$ that models\linebreak
\vspace{62.5pt}\newline
losses in the circuit.
The inductance of the bare resonator is perturbed by the magnetic susceptibility of the sample
\begin{equation}
\chi=\chi'-i\chi'',
\end{equation}
such that the resonator inductance is given by \begin{equation}L = L_0 \left[ 1+ \eta \chi' -i \eta \chi''\right],\end{equation} where the inductance of the bare loop is given by $L_0$ and the filling factor $\eta$ parameterises the overlap between the sample and the oscillating magnetic field of amplitude $B_1$ and angular frequency $\omega$ \cite{book:Slichter1990}. By considering the impedance \begin{equation}Z=\left[ i \omega C + \frac{1}{i \omega L + R}\right]^{-1}\end{equation} of the resonant circuit, and by treating the effect of the sample magnetic susceptibility as a perturbation, such that $\eta\chi(\omega)\ll 1$, one can derive the expression \begin{equation}
\omega_\mathrm{res}=\omega_0\left[1- \frac{1}{2} \eta \chi'\right],
\label{equation:frequency}
\end{equation} where 
\begin{equation}\omega_0=\frac{1}{\sqrt{L_0 C}} \left( 1- \frac{CR^2}{2L}\right)\end{equation}
is the resonant frequency of the bare resonator. These calculations assume that the quality factor of the resonator $Q\gg1$.

In the absence of saturation, the value of $\chi'$ is given by \begin{equation}
\chi'(\omega_\mathrm{spin},\omega_\mathrm{rad})=\overline{\chi}(\omega_\mathrm{spin}) \frac{(\omega_\mathrm{spin}-\omega_\mathrm{rad})}{(\delta \omega /2)^2+(\omega_\mathrm{rad}-\omega_\mathrm{spin})^2 }, 
\label{equation:chiprime}
\end{equation} where $\omega_\mathrm{spin}$ is the resonant frequency of the spin ensemble at a field $B_0$, $\omega_\mathrm{rad}$ is the frequency of applied radiation, and $\delta \omega$ is the width of the resonance in the angular frequency domain for a given value of $\omega_\mathrm{rad}$. The quantity $\overline{\chi}$, with dimension $\si{\second}^{-1}$, parameterises the magnitude of the magnetic susceptibility for a given energy splitting between the clock states.

In general, $\omega_\mathrm{spin}$ is a non-linear function of magnetic field $B_0$; however, over the small range of magnetic field where the ensemble is near resonance ($\omega_\mathrm{spin}\approx\omega_\mathrm{rad}$), we can approximate it as a linear function such that $\omega_\mathrm{spin}-\omega_\mathrm{rad}=\gamma_\mathrm{eff}\Delta B$, where $\gamma_\mathrm{eff}$ is the effective gyromagnetic ratio at the predicted resonant field $B_\mathrm{rad}$ for which $\omega_\mathrm{spin}=\omega_\mathrm{rad}$, and $\Delta B=B_0-B_\mathrm{rad}$. Furthermore, we take the susceptibility parameter $\overline{\chi}$ as constant over one linewidth, such that $\overline{\chi}(\omega_\mathrm{spin})\approx \overline{\chi}(\omega_\mathrm{rad})$. In this case, Equation (\ref{equation:chiprime}) reduces to \begin{equation}
\chi'(\Delta B,\omega_\mathrm{rad})=\frac{\overline{\chi}(\omega_\mathrm{rad})}{\gamma_\mathrm{eff}}\cdot \frac{( \Delta B) }{( \delta B / 2)^2+(\Delta B)^2 },
\label{equation:chiprimefield}
\end{equation} where the field domain linewidth $\delta B = \delta \omega / \gamma_\mathrm{eff}$. Substituting Equation (\ref{equation:chiprimefield}) into Equation (\ref{equation:frequency}), we can see that the frequency shift $\Delta \omega_\mathrm{res}$ of the resonator is given by the derivative of a Lorentzian in the field domain. For small cavity frequency shifts, the reflection of the detuned excitation radiation is linearly converted into an amplitude signal by the frequency-dependent reflection coefficient of the resonator.
The reflected signal amplitude is therefore described by a Lorentzian-derivative lineshape $S\propto \chi'(\Delta B,\omega_\mathrm{rad})$, and hence the observed signal after field modulation and phase-sensitive detection \begin{equation}S_\mathrm{obs} = k \frac{\partial}{\partial (\Delta B)} \bigg[\chi'(\Delta B,\omega_\mathrm{rad})\bigg] \label{supeqn:obs_signal}\end{equation} is described the second-derivative of a Lorentzian in the field domain. The experimental parameter $k$ parameterises the spectrometer's sensitivity. The resonance signals in each fixed-frequency, swept-field spectrum are fitted using an equation whose functional form is given by Equation (\ref{supeqn:obs_signal}), with the addition of a dc offset due to, for example, inductive pickup of the modulation field. The fit parameters are signal amplitude $S_0$,  field domain linewidth $\delta B$, resonant field $B_\mathrm{res}$, and dc~offset.

\label{supsec:spectrumfit}
\section{Fitting the resonance locations}

To fit the data shown in Figure 3 and Figure 4 of the main text, the spin Hamiltonian was diagonalised to give analytic expressions for the energy eigenlevels $E_1(B_0)\ldots E_8(B_0)$ as a function of magnetic field $B_0$. From these expressions, we calculate the transition frequencies $f_1 \ldots f_N$ as a function of magnetic field. The set of resonant fields derived from the fits described in Section \ref{supsec:spectrumfit} was divided into subsets $i$, each corresponding to a particular allowed transition frequency $f_i=f_i(B_0,A,g_\mathrm{e},g_\mathrm{N})$.

We find it necessary to account for small differences between the nominal field $B_\mathrm{coil}$ generated by the Helmholtz coil and the true field $B_0$ at the sample by fitting the data using $B_0=\alpha B_\mathrm{coil} + B_\mathrm{offset}$, where $\alpha$ is a correction parameter, reflecting sample misalignment and uncertainty in the calculated coil constant, and $B_\mathrm{offset}$ is an environmental offset field. Therefore, we fit the set of resonant fields for transition $i$ at a given transition frequency using the relationship $f_i=f_i(B_\mathrm{coil},B_\mathrm{offset},\alpha,A,g_\mathrm{e},g_\mathrm{N})$, using $B_\mathrm{offset}$, $\alpha$ and $A$ as the fit parameters. We can fit multiple transitions simultaneously, as in Fig.~3, or fit a single transition, as in Fig.~4, using the \texttt{lsqnonlin} function provided by the \textsc{Matlab} Optimization Toolbox. In both cases, the extracted $\alpha$ and $B_\mathrm{offset}$ are used to plot the data in terms of $B_0$ rather than $B_\mathrm{applied}$. The one standard deviation error bars quoted are calculated from $\sigma=\Delta \mathrm{C.I.}/4,$ where $\Delta \mathrm{C.I.}$ is the difference between the upper- and lower-value of the fitted 95\% confidence intervals.

We find good agreement between the values of $A$ and $\alpha$ determined by fitting simultaneously the parallel and perpendicular datasets and the values determined by fitting solely to the clock transition dataset. However, we observe a small difference between the fitted value of the offset field, which is $B_\mathrm{offset}=-28\pm4\,\si{\micro\tesla}$ for the former case and $B_\mathrm{offset}=-12.1\pm0.5\,\si{\micro\tesla}$ for the latter case. This difference between the offset fields is physically reasonable because the energy splitting is determined by $B_0=|\alpha\boldsymbol{B}_{\mathrm{coil}}+\boldsymbol{B}_{\mathrm{env}}|$, where $\boldsymbol{B}_\mathrm{env}$ is a fixed environmental field set by, for example, the Earth's magnetic field. The fitted value of $B_\mathrm{offset}$ therefore depends on the relative orientation of the magnet field $\boldsymbol{B}_\mathrm{coil}$ and $\boldsymbol{B}_\mathrm{env}$, and hence differs between perpendicular and parallel mode operation.

\end{document}